\documentclass[prl,preprint,aps,epsf]{revtex4}
\usepackage{graphics}
\usepackage{epsfig}
\begin{document}
\draft
\title{Non-Reversible Evolution of Quantum Chaotic System. Kinetic Description.}
\author{L. Chotorlishvili, V. Skrinnikov }
\affiliation{Department of Physics,Tbilisi State University, Chavchavadze av. 3, 0128 Tbilisi, Georgia \\
E-mail: lchotor33@yahoo.com}
\date{\today}

\begin{abstract}
It is well known that the appearance of non-reversibility in classical chaotic systems is connected with a local instability of phase trajectories relatively to a small change of initial conditions and parameters of the system. Classical chaotic systems reveal an exponential sensitivity to these changes. This leads to an exponential growth of the initial error with time, and as the result after the statistical averaging over this error, the dynamics of the system becomes non-reversible.

In spite of this, the question about the origin of non-reversibility in quantum case remains actual. The point is that the classical notion of instability of phase trajectories loses its sense during quantum consideration.

The current work is dedicated to the clarification of the origin of non-reversibility in quantum chaotic systems.

For this purpose we study a non-stationary dynamics of the chaotic quantum system. By analogy with classical chaos, we consider an influence of a small unavoidable error of the parameter of the system on the non-reversibility of the dynamics.

It is shown in the paper that due to the peculiarity of chaotic quantum systems, the statistical averaging over the small unavoidable error leads to the non-reversible transition from the pure state into the mixed one.

The second part of the paper is dedicated to the kinematic description of the chaotic quantum-mechanical system. Using the formalism of superoperators, a muster kinematic equation for chaotic quantum system was obtained from Liouville equation under a strict mathematical consideration.
\end{abstract}


\maketitle

\section*{INTRODUCTION}

Study of the quantum reversibility and motional stability is of
great interest [1-4]. This interest is due to not only the
fundamental  problem of irreversibility in quantum dynamics, but
also to particular application. In particular, it reveals itself in relarion to the
field of quantum computation. A quantity of central importance
which has been on the focus of many studies [5-9] is the so-called
fidelity $f(t)$, which measures the accuracy to which  a quantum
state can be recovered by inverting, at  time $t$, the dynamics
with a perturbed Hamiltonian
\begin{equation}
f(t)=|\langle \psi|e^{i\hat{H}t}e^{-i\hat{H}_ot}|\psi\rangle|^2.
\end{equation}

Here $\psi$ is the initial state which evolves in time $t$ with
the Hamiltonian $\hat{H}_o$, while $\hat{H}=\hat{H}_o+\hat{V}$ is
the perturbed Hamiltonian. The analysis of this quantity has shown
that under some restrictions, the series taken from $f(t)$ is exponential
with a rate given by the classical Lyapunov  exponent [5-6]. But
here a question appears. The point is that the origin of the dynamic
stochasticity, which is the reason of irreversibility in
classical case, is directly related to the nonlinearity of
equation of motion. For classical chaotic system this nonlinearity
leads to the repulsion of phase trajectories at a sufficiently
quick rate [10-13]. In case of quantum consideration, the
dynamics of a system is described by a wave function that obeys a
linear equation and the notion of a trajectory is not used at all.
Hence, at first sight it seems problematic to find out the quantum
properties of systems, whose classical consideration reveals their
dynamic stochasticity. For this reason, according to the widely
accepted opinion the notion of "Quantum Chaos" includes phenomena,
related to the quantum-mechanical description of systems chaotic
in the classical limit [14-17].

Unfortunately, our understanding of quantum dynamics of chaotic
systems is still quite limited. The majority of the existing
quantum chaos literature concentrates on understanding the
properties of eigenfunctions and eigenvalues.

In this paper we consider the issue of irreversibility for
quantized chaotic system. For simplicity we shall discuss a
periodic driving.

In given work we shall try to show that in essentially
chaotic domain, quantum chaotic dynamics is characterized by the
transition from the pure quantum-mechanical state into the mixed
one. In this case, irreversibility in the system appears as a
result of loss of informaintion about the phase factor of the wave
function. To prove this phenomenon, we shall consider model
Hamiltonian studied in [1,18]. We shall try to give our
explanation of the mixed state formation in a chaotic
quantum-mechanical system. The main purpose of given work is the
deriving of kinetic equation for chaotic quantum-mechanical
system. Interesting method of how to derive a kinetic equation for
semi-classical chaotic system was offered in [16]. In this work we
shall try to do the same in exceptionally quantum domain, without
application to the semi-classical methods. The paper is organized
as follows. In the first part formation of a mixed state is
considered. The second part is dedicated to the kinetic
description of quantum chaos.

\section{Formation of mixed state.}

During the discussion of nonreversible evolution of the
quantum-mechanical system naturally the question emerges. How  does a non-reversibility originate in a
quantum system? If the quantum system
evolves according to the Schrodinger equation, how can a pure
quantum-mechanical state become a mixed one? The question is that,
in contrast to the classical chaos, quantum-mechanically irregular
motion cannot be characterized by extreme sensitivity to tiny
changes of the initial data. Due to the unitarity of the quantum
dynamics, the overlap of two wave functions remains
time-independent
$|\langle\Theta(t)|\xi(t)\rangle|^2=|\langle\Theta(0)|\xi(0)\rangle|^2$,
provided time-dependence of $\Theta(t)$ and $\xi(t)$ is generated
by the same Hamiltonian. However, an alternative characterization
of classical chaos, extreme sensitivity to slight changes of the
dynamics does carry over into quantum mechanics.

As it was mentioned in the introduction, the model Hamiltonian,
studied in [1,18] is taken as a model for our considerations

\begin{equation}
\hat{H}(Q,P,x)=\frac{1}{2}(P^2_1+P^2_2+Q^2_1+Q^2_2)+xQ^2_1Q^2_2,
\label{2}
\end{equation}
where $P_1$, $P_2$, and $Q_1$, $Q_2$ are canonical momenta and
coordinates. Here and further we shall consider dimensionless
quantities.

It follows from works [1,18] that for values of parameters
$x=x_o\approx 1$ and $E\approx 3$ (where $E$ is energy of system),
dynamics related to Hamiltonian (2) is chaotic and is
characterized by correlation function, the width of which is equal to
$\tau_c=1$ [1]. Our goal is to study mechanisms of irreversibility and mixed state formation in non-stationary quantum dynamics. For this let present Hamiltonian (2) in the following
form $$\hat{H}(t)=\hat{H}_o+V(t),$$
\begin{equation}
\hat{H}_o=\frac{1}{2}(P^2_1+P^2_2+Q^2_1+Q^2_2)+x_oQ^2_1Q^2_2,
\end{equation}
$$V(t)=V_o(Q)f_o(t),~V_o=\Delta x_o \cdot Q^2_1Q^2_2.$$

As one can see from (3), the time dependence of the Hamiltonian
$\hat{H}(t)$ is provide by amplitude modulation of the parameter
$x=x_o+\Delta x_of_o(t)$, where $f_o(t)$ is a periodic function
of time with $T_o$ period. Hereinafter we shall work in a
domain of chaotic motion  $x_o\sim 1,~ \Delta x_o<x_o$.

After taking (3) into account, the Schrodinger equation for the
wave function takes the form
\begin{equation}
i\frac{\partial |\psi(Q,x(t))\rangle}{\partial
t}=\hat{H}(Q,x(t))|\psi(Q,x(t))\rangle. \label{4}
\end{equation}

The solution of the time-dependent  Schrodinger equation (4)
 can be written formally with the help a time-dependent
exponential [14]
$$U(t)=exp[-i\int_{o}^{t}dt'\hat{H}(t')]_t,$$
where the positive time ordering requires $$[A(t)B(t')]_t=
\left\{
\begin{array}{l}
A(t)B(t')\hskip 0.7cm{\mbox if}\hskip 0.2cm t>t' \\
B(t')A(t)\hskip 0.7cm {\mbox if} \hskip 0.2cm t<t'
\end {array} .
\right. $$

In our case $H(t+T_ok)=H(t)$, $k=1,2...$, the evolution operator
referring to one period $T_o$ ,  the so-called Flouqet operator
$U(T_o)\equiv \hat{F}$ [14], is worthy of consideration, since it
yields a stroboscope view of the dynamics
$$|\psi(kT_o)\rangle=(\hat{F})^n |\psi(0)\rangle.$$

The Flouqet operator being unitary has unimodular eigenvalues.
Suppose we can find eigenvectors $|\varphi _{\chi}\rangle$ of the
Flouqet operator $$\hat{F}|\varphi _{\chi}\rangle =
e^{-i\varphi_{\chi}}|\varphi _{\chi}\rangle,$$
\begin{equation}
\langle \varphi _{\chi}|\varphi _{\Theta}\rangle =\delta _
{\chi\Theta}. \label{5}
\end{equation}
Then, with the eigenvalue problem solved, the stroboscopic
dynamics may be written explicitly [14]
\begin{equation}
|\psi_{n}(kT_o)\rangle = \sum_{\chi} e^{-ik \varphi_{\chi}}\langle
\varphi _{\chi}|\psi_{n}(0)\rangle|\varphi _{\chi}\rangle.   \label{6}
\end{equation}

Here $|\psi_{n}(0)\rangle$ is the eigenfunction of the Hamiltonian
$H$ (3), corresponding to the value of parameter $x(0)=x_0+\Delta
x_0f(0)$.

 As it was mentioned above, our aim is to prove that one of the signs of the emergence
of quantum chaos is a formation of the mixed state. Being
initially in a pure quantum-mechanical state, described by the
wave function $|\psi_n\rangle$, the system during the evolution
makes an irreversible transition to the mixed state.

The information about whether the system is in the mixed state or in the pure one, may be obtained from the form of the density matrix [19-21]. Introducing the definition $C_{nm}(k)=e^{-ik\phi_m}\langle\phi_m | \psi_n(0)\rangle$, let us rewrite (6) in more convenient way for further usage:
\begin{equation}
| \psi_{n}(kT_0)\rangle = \sum_{m}C_{nm}(k)| \phi_m\rangle.
\end{equation}
Next, according to the commonly accepted rules [22], the density matrix of the system may be expressed via the expansion coeficients $C_{nm}(k)$
\begin{equation}
\rho_{nm}(k)=\sum_{p}C_{np}(k)C_{mp}^*(k).
\end{equation}
After the substitution of the explicit form of $C_{nm}(k)$ coefficients into (8), we get:
\begin{equation}
\rho_{nm}(k)=A_{mn}e^{ik(\phi_m-\phi_n)},
\end{equation}
where
\begin{equation}
A_{mn}=\sum_p \langle\phi_m | \psi_p(0)\rangle\langle\psi_p(0) | \phi_n\rangle.
\end{equation}

As seen from (9), non-diagonal matrix elements of the density matrix contain fast oscillating in time exponential phase factors $e^{ik(\phi_m-\phi_n)}$, and the diagonal matrix elements
\begin{equation}
\rho_{nn}=A_{nn}=\sum_p |\langle\phi_n | \psi_p(0)\rangle |^2.
\end{equation}

Exponential phase factors of the non-diagonal matrix elements $\rho_{nm}(k)$ express the principle of quantum coherence [23] and correspond to the complete quantum-mechanical description of the system in pure quantum-mechanical state. While they are not equal to zero, the system is in the pure state, and zeroing of these elements is the sign of transition into mixed state [20-22].

But here a question appears: if the system was initially in the pure quantum-mechanical state, how can the transition into mixed state take place? Before answering on this question, let us recall, that existence of non-complete quantum information about the state of the system corresponds to the mixed quantum-mechanical state [19-23]. I.e. for transition into the mixed state, a partial loss of information about the state of the system must occur. According to the commonly accepted notions [20], this corresponds to the loss of information about the phases of the system and to the zeroing of non-diagonal elements of the density matrix.

So, to prove the formation of the mixed state one has to show
the zeroing of non-diagonal elements of density matrix, the
equality of which to zero is a sign of a mixed state [19-21].

Let us recollect that value $\phi_n$ is the
eigenvalue of the Floquet operator. Owing to the non-integrability
of the system, eigenvalues can be obtained only by way of
numerical diagonalization of the Hamiltonian $H_0$ (3). According
to the main hypothesis of the random matrix theory [14,17],
the elements of this matrix in the chaotic domain are random numbers.
So, their eigenvalues can also be considered as random values.

This statement is valid only in the chaotic domain [24,25].
According to this, the phase
\begin{equation}
f(n,m)=\phi_m-\phi_m,
\end{equation}
in exponential factors of the non-diagonal matrix elements in (9)
is random quantity with all its properties. Such an interpretation
of the eigenvalues of chaotic quantum-mechanical system is stated
in recently published works [24,25]. In these papers the analogy
between spectral characteristics of a Hamilton system and random
time series is discussed. Further we shall treat $f(n,m)$ phase as
a mathematical random variable putting aside its physical meaning.
According to the preceding, it is clear that the values of matrix
elements of density matrix of the chaotic quantum-mechanical
system $\rho_{nm}(k)$ are random values too. Taking a statistical
average of expression (9), we have:
\begin{equation}
\langle \rho_{nm}(k)\rangle = \langle A_{mn}e^{ik(\phi_m -
\phi_n)}\rangle ,
\end{equation}
where $\langle ... \rangle$ denotes the averaging over the statistical ensemble.

Before proceeding the discussion, we should define what one implies under quantum-statistical ensemble in our case more precisely.

Recall, that while studying classical chaos, one usually examines the stability of the system relatively to a small change of initial conditions and system parameters. A small initial error of these parameters always exists and remains unavoidable (one may measure the parameters of the system at a very high precision, but even in this case there is still a small error, the removal of which, i.e. the measurement at absolute precision, is impossible [10]). So that, not the existence of the unavoidable error is fundamental, but what kind of influence it brings over the system dynamics. It is well known that in case of regular systems such an influence is negligible, but if the system is chaotic, the effect of it increases exponentially [12].

Making a complete analogy to classical chaos, let us assume, that the parameter, characterizing interaction between oscillators $x_0$, has some small unavoidable error $\delta x_0$. Then the set of chaotic Hamiltonians (3), with the values of interaction parameter taken from interval $[x_0-\delta x_0, x_0+\delta x_0]$, may be considered as quantum-mechanical ensemble. According to the random matrix theory [14,17,27], there is a random distribution of distances between levels that corresponds to ensemble of chaotic Hamiltonians. In particular for chaotic Hamiltonians with real matrix elements the distribution of the distances between levels obeys normal Gaussian distribution [14,17].

The further study of (13) can be performed using more rigorous mathematical
substantiation. For this let us recall some details from the
probability theory [26].

a) In general case, under the characteristic function of the
random variable $X$, mathematical expectation of the following
exponent is meant
\begin{equation}
F(t)=M(exp(itX)),
\end{equation}
where $t$ is a real parameter.

b) Mathematical expectation itself is defined as a first initial
moment $\mu_1$ of the random variable $X$
\begin{equation}
M(X)=\langle X\rangle =\mu_1=\sum_k x_kP_k,
\end{equation}
where $X$ is the discrete random variable which takes possible
values $x_1,x_2,...$  with appropriate probabilities
$P_k=P(X=x_k), <...>$ means average.

Taking (14), (15) into account and considering $f(n,m)$ as a
random value, we get:
\begin{equation}
\langle\rho_{nm}(k)\rangle=A_{mn}F(k),
\end{equation}
where
\begin{equation}
F(k)=\langle e^{ikf(n,m)}\rangle,
\end{equation}
is the mathematical expectation of the characteristic function of
the random phase $f(n,m)$, and
\begin{equation}
A_{mn}=\langle\phi_m|\psi(0)\rangle\langle\psi(0)|\phi_n\rangle.
\end{equation}

Let us assume that random phase $f(n,m)$ has a normal dispersion
[26]. This assumption is based on the fact that matrix elements of Hamiltonian (3) are real values. Then from (17) one can obtain
\begin{equation}
F(k)=e^{iak}e^{-\frac{\sigma^2k^2}{2}},
\end{equation}
where
\begin{equation}
a=M(f(n,m))
\end{equation}
is the mathematical expectation of the random phase,
\begin{equation}
\sigma^2=M(f^2(n,m))-(M(f(n,m)))^2.
\end{equation}

On basis of obtained expressions (16), (19) one may say the following. For small $k$, the system is still in a pure state (non-diagonal matrix elements differ from zero). But after some time the system passes to the mixed state as the non-diagonal matrix elements of the density matrix decrease with exponential law
\begin{equation}
\rho_{nm}(k)\sim e^{-\frac{\sigma^2k^2}{2}}.
\end{equation}
This phenomenon is connected with the "phase incursion" [12]. Uncertainty of phase in (9) is accumulated little by little with time. Finally, when $k>\sqrt{2}/\sigma$, the uncertainty of phase is of order $~2\pi$, and the phase is totally chaotized (see [3,4]). As the result the system passes into mixed state.

But here a question emerges. How does the uniqueness of the nonlinear chaotic system show itself? If we assume the existence of initial dispersion of parameter for linear system, shall we obtain the same result? Before answering on this question let us consider a peculiarity of already obtained result (22):

When approaching to (22), the fact of randomicity of the distribution of distances between levels of energy terms was fundamental. We should notice, that this randomicity was made conditional on the randomicity of the distribution of energy termes themselves and not due to the widening of the line $\delta\phi_n$ of the energy terms $\phi_n$ [14,17].

Now let us consider a linear system and see whether this condition is carried out or not.

Consider ensemble of linear oscillators
\begin{equation}
H_L=\frac{p^2}{2m}+\frac{\beta x^2}{2},
\end{equation}
that are characterized by a small dispersion of parameter $\delta\beta\ll\beta$.

The spectrum of linear oscillator is well known and has a regular character [20]
\begin{equation}
E_n(\omega)=\omega(n+\frac{1}{2}),
\end{equation}
where $\omega=\sqrt{\beta/m}$ is the distance between the levels of the system.

Easy to see, that if uncertainty of parameter $\beta$ is small $\delta\beta\ll\beta$, then this may lead to the widening of the levels, but not to the mixing of them. As a result, the value of such widening in $\omega$ units is:
\begin{equation}
\delta E_n\approx\delta\omega(n+\frac{1}{2}),
\end{equation}
where $\delta\omega\Rightarrow\frac{\delta\omega}{\omega}=\frac{1}{2}\delta\beta$.

It is already clear, that the condition is not fulfilled. In this case the damping of non-diagonal matrix elements of density matrix may be caused only by the widening of energy terms and not by their random distribution. One should take into account that the widening of term depends linearly on the dispersion $\delta\beta$, and each of the Hamiltonians makes an equal contribution to the average value. This happens because in the case of linear oscillator, each of the Hamiltonians of the ensemble has the same spectrum as others have, and the averaging is performed over the widening of energy terms. In case of chaotic ensemble, each of the Hamiltonians of the system has its own spectrum (due to exponential repulsion of the energy terms of the chaotic ensemble [14,17] they are exponentially sensitive to a small dispersion) and the averaging is performed over these spectrum characteristics, and not over the widening of separate line. When taking such average, the contribution from separate Hamiltonians of the chaotic ensemble is not the same and is described by Gaussian esemble [14,17]. This is the fundamental and not the formal difference between these two cases.

Taking all this into account, we obtain for ensemble average matrix elements of linear oscillator:
\begin{equation}
\langle\rho_{mn}^L(t)\rangle=\frac{1}{2\delta\beta}\int_{-\delta\beta}^{\delta\beta}\rho_{mn}^L(t,\delta\beta')d(\delta\beta').
\end{equation}

Considering (24), (25) and (26) we get in the end:
\begin{equation}
\langle\rho_{mn}^L(t)\rangle_{\delta\beta}=\frac{2}{\delta\beta t|n-m|}e^{i(m-n)\omega t}\cos{\left( \frac{\delta\beta}{2}(n-m)\omega t\right)}.
\end{equation}

Let us start the analysis of the obtained result. Firstly it is clearly seen, that time dependence is not of Gaussian type, but is inverse proportional function of time $1/t$. This as minimum decreases the rate of formation of mixed state in comparison to the case of chaotical ensemble (see Eq.(22)). More than this, the time of formation of the mixed state is completely defined by the value of dispersion $\delta\beta$:
\begin{equation}T_c=\frac{2}{|n-m|}\frac{1}{\delta\beta}.
\end{equation}

Actually this means the following. Observing the evolution of the system on an arbitrary large interval of time $t\in[0,T]$, one can always select such an accuracy while measuring the system parameter $\delta\beta_0$ to fulfill the condition $T<T_c$. This means that the system will be in a pure state for given observation interval $t\in[0,T]$ and given accuracy of measurement $\delta\beta_0$.

Hence in some sense we have an analogy with classical chaos. A small initial dispersion of values has an exponential effect, which reveals itself in formation of mixed state. In case of regular system, one may neglect this dispersion with an accurace up to the widening of the energy terms.

Finally let us discuss one more question. Can one formally obtain the result (22) in a case of linear oscillator? From a formal mathematical point of view it is possible, but this case will not have a physical sense. For this one has to choose an ensemble of linear oscillators with predetermined random frequences that obey Gaussian distribution. Next one does the averaging over this ensemble, taken over the random frequences and not over the widening of lines. But this case will not have anything common with a real physical situation. The thing is that the Gaussian ensemble of Hamiltonians in case of chaotic system as well as the ensemble of harmonic oscillators in a case of regular system is directly connected with an unavoidable error of measurement of physical parameters of the system, and the averaging over them is not a pure formality.

Zeroing of the non-diagonal elements of density matrix is the sign
of quantum chaos beginnings and formation of mixed state. From
that moment the quantum dynamics is non-reversible, since the
information about the  wave functions phase is lost.

After the formation of a mixed state the quantum-mechanical
description loses its sense and we have to use a
quantum-statistical interpretation. Along with this, the notion of
fidelity loses its meaning, i.e. by the time of the beginning of
reverse evolution the system is no longer in the pure state
described by the wave-function. The reverse transition from the
mixed state to the pure one and reconstruction of the
wave-function are impossible because of the irreversibility of the
process.

Given above reasoning can be checked by numerical diagonalization of the
Hamiltonian (2), defining eigenvalues $\varphi_{\chi}$ and by
estimation of the averaged non-diagonal elements of the density
matrix according to (9).

The result of numerical calculations is represented on Fig.1.

\begin{figure}[tbp]
\includegraphics[scale=0.5]{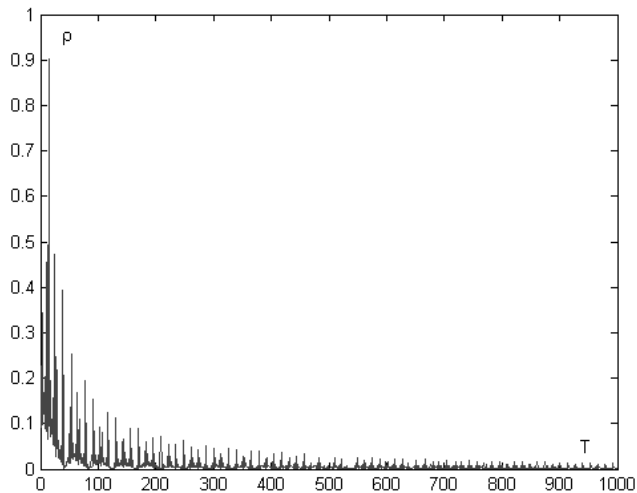}
\caption{
\label{Fig.1} The graph of the dependence $\rho_{nm}$  (9) on time
$t=kT_o$ . The graph is obtained using (9) and by way of numerical
diagonalization of the Hamiltonian (3) $\hat{H}(P,Q,x(0))$. As it
is easily seen from the graph that when $t=kT_o=\sqrt{2}/\sigma>\tau_c,~
T_o=\frac{2\pi}{\Omega}=0.1\tau_c$, $\tau_c=1$ [1], zeroing of
$\rho_{nm}$ happens. In the numerical calculations we have
used definite initial state $\psi(0)$, where $\psi(0)$ is the
eigenvector of the Hamiltonian $\hat{H}(P,Q,x(0)), x(o)=x_o+\Delta
x_o$, and definite eigenfunctions $|\phi_n\rangle$,
$|\phi_m\rangle$, $n=6$, $m=80$ of the Hamiltonian
$\hat{H}(P,Q,x_0)$. The dimension of the diagonalized matrix was
$1000 \times 1000,\Delta x_o=0.1$. For the statistical averaging
of the numerical data we have used ensemble of 50 Hamiltonians,
corresponding to the small dispersion of parameter $\delta
x_0=0.01$.
}

\end{figure}

\section{Kinetic description of the chaotic quantum-mechanical
system.}

In previous section we studied mechanism of mixed state
formation. After formation of mixed state quantum-mechanical
consideration loses meaning and there is a need to use a kinetic
description. Kinetic equation for the chaotic quantum-mechanical
system for the first time was obtained in [16]. But this study was done in
the semi-classical domain. Namely for example, zeroing of
non-diagonal part of density matrix was proved by use of
semi-classical approximation. The main purpose of given work is
the deriving of kinetic equation for chaotic quantum-mechanical
system in exceptionally quantum domain without application to the
semi-classical methods. For obtaining of muster equation for the
density matrix we shall use a method of projection operator [28].

Let us split the density matrix operator $\hat{\rho}$ on a slow
$\hat{\rho}_{R}$ and a fast varying $\hat{\rho}_{NR}$ operators:
$\hat{\rho}=\hat{\rho}_{R}+\hat{\rho}_{NR}$. Relevant part
$\hat{\rho}_{R}$ in the basis $|\varphi_n \rangle$ contains only
diagonal elements, whereas non-relevant part $\hat{\rho}_{NR}$
contains only non-diagonal elements. These elements, as it was shown in previuos section, contain fast oscillating exponents and when taking average over the ensemble, the zeroing of them takes place.  Elimination of the diagonal
part from the density matrix is a linear operation, which
satisfies the property of projection operator $\hat{D}^2=\hat{D}$
[28]
\begin{equation}
\hat{\rho}_{R}=\hat{D}\hat{\rho},~~
\hat{\rho}_{NR}=(1-\hat{D})\hat{\rho}.
\end{equation}

Let us note that this reflection is non-reversible. Due to the
zeroing of non-diagonal part of the density matrix, part of
information is lost.

Inasmuch as  relevant statistical operator $\hat{\rho}_{R}(t)$ is
different from the total operator $\hat{\rho}(t)$, generally
speaking it does not suit the Liouville-Fon Neumann equation [21]

\begin{equation}
\frac{\partial \hat{\rho}}{\partial
t}+i\hat{L}\hat{\rho}=0,
\end{equation}
where $\hat{L}$ is Liouville operator [19]. After acting on the
equation (30) by operator $\hat{D}$, we get
 $$\frac{\partial
\hat{\rho} _{R}}{\partial
t}+i\hat{D}\hat{L}(\hat{\rho}_{R}+\hat{\rho}_{NR})=0,$$

\begin{equation}
\frac{\partial \hat{\rho} _{NR}}{\partial
t}+i(1-\hat{D})\hat{L}(\hat{\rho}_{R}+\hat{\rho}_{NR})=0.
\end{equation}

For the purpose to obtain closed equation for $\hat{\rho} _{R}$ we
exclude  from the first equation (31) $\hat{\rho} _{NR}$ . As a
result we get $$\frac{\partial \hat{\rho} _{R}(t)}{\partial
t}+i\hat{D}\hat{L}\hat{\rho}_{R}(t)+\int\limits_{t_o}^{t}
K(t-t_1)\hat{\rho}_{R}(t_1)dt_1=-i\hat{D}\hat{L}exp[-i(t-t_o)(1-\hat{D})\hat{L}]\hat{\rho}_{NR}(t_o),$$
\begin{equation}
K(t-t_1)=\hat{D}\hat{L}exp[-i(t-t_1)(1-\hat{D})\hat{L}](1-\hat{D})\hat{L}.
\end{equation}
This equation is valid for $t=\sqrt{2}/\sigma>\tau_c$. Evidently
$\hat{\rho}_{R}(t)$ is expressed by way of values of
$\hat{\rho}_{R}(t_1)$ taken for the time interval $t_o<t_1\leq t$,
and additionally trough the value of $\hat{\rho}_{NR}(t_o)$. If in
the initial moment of time $t=t_o$ the system is in a pure
quantum-mechanical state then $\hat{\rho}_{NR}(0)\not=0$.

In this case solving the equation (32) is problematically. However
let us recollect that for $t_o>\sqrt{2}/\sigma$  system is already in a
mixed state. Therefore equation (32) takes more simple form
$(\hat{\rho}_{NR}(t_o)=0)$
\begin{equation}
\frac{\partial \hat{\rho} _{R}(t)}{\partial
t}+i\hat{D}\hat{L}\hat{\rho}_{R}(t)+\int\limits_{t_o}^{t}
K(t-t_1)\hat{\rho}_{R}(t_1)dt_1=0.
\end{equation}

For solving equation (33) we shall use a method of super-operators
[29].  But before performing this, we should notice, that we have obtained a closed equation for relevant part of statistical operator. We were able to come to this only because when $t>t_0=\sqrt{2}/\sigma$ the system is in the mixed state and all non-diagonal matrix elements of the density matrix are equal to zero.

Further when studying the evolution of the system we shall consider as the origin of time the moment of the formation of the mixed state in the system. This corresponds to a formal transition to limit $t_0\rightarrow -\infty$. Further for simplification of (33) we shall use Abel theorem [26]
\begin{equation}
\lim_{T\rightarrow \infty}\frac{1}{T}\int_T^0 f(t)dt=f(0)-\lim_{\epsilon\rightarrow +0}\int_{-\infty}^{0}e^{\epsilon(t')}\frac{d}{dt'}f'(t')dt' .
\end{equation}
Taking (34) into account, Eq.(33) will have the following form:
\begin{equation}
\frac{\partial\hat{\rho}_R(t)}{\partial t}+iDL\hat{\rho}_R(t)=\lim_{\epsilon\rightarrow +0}\int_{-\infty}^{t}e^{\epsilon (t' -t)}K(t-t')\hat{\rho}_R(t')dt' .
\end{equation}
According to the method of superoperators [29], the correspondence of one operator to another may be considered as representation. The operator in this case will be represented by a matrix element with two indices, while the linear product of operators is a matrix with four indices, i.e. a superoperator [29]. A concrete example is the projection of $\hat{D}$ operator on diagonal elements. We should notice, that in our case, the projection operator is some definite physical procedure of averaging matrix elements of the density matrix over Gaussian chaotic ensemble.

From the relation $\hat{D}\rho_{nm}(t)=\rho_{nm}(t)\delta_{mn}$ we come to the following representation of $\hat{D}$ superoperator $\hat{D}_{nmn'm'}=\delta_{nn'}\delta_{mm'}\delta_{nm}$, so that
\begin{equation}
\sum_{m'n'}\hat{D}_{mnm'n'}\rho_{m'n'}=\rho_{nm}\delta_{nm}.
\end{equation}
Taking (36) into account, Eq.(35) is
\begin{equation}
\frac{d\rho_{nn}(t)}{dt}+i\hat{D}(\hat{L}\hat{\rho}_R)_{nn}=-\int_{-\infty}^{0}dt'[K(-t')\hat{\rho}_R(t+t')]_{nn}e^{\epsilon t'}.
\end{equation}
In expression (37) and further for short we shall omit $R$ index for diagonal matrix elements of the operator $\hat{\rho}$. Considering the relation
\begin{equation}
[\hat{L},\hat{\rho}_R]_{nn} = \sum_a L_{nnaa}\rho_{aa}=0,
\end{equation}
and representing Liouville operator as $L=L_0+L'$ in compliance with (3) we get:
\begin{equation}
\frac{d\rho_{nn}(t)}{dt}=-\int_{-\infty}^{t}dt_1 e^{\epsilon(t-t_1)}\sum_{n}K_{nnmm}(t-t_1)\rho_{mn}(t_1),
\end{equation}
and
\begin{equation}
K_{nnmm}(t)=[L' e^{-it(1-D)L}(1-D)L']_{nnmm}.
\end{equation}
From (40), (3) and from representation of Liouville operator in form $L=L_0+L'$, one can see, that kernel $K_{nnmm}(t)$ is at least of second order by $\Delta x_0$. Next it is easy to check the corectness of the expression:
\begin{equation}
\sum_{m}L_{abmm}'=\sum_m(V_{am}\delta_{bm}-V_{mb}\delta_{am})=V_{ab}-V_{ab}=0,
\end{equation}
for Liouville superoperator
\begin{equation}
L_{mnm'n'}=(H_{mm'}\delta_{nn'}-H_{nn'}\delta_{mm'}).
\end{equation}
The relation (41) in its turn leads to the rule of sums $\sum_{m}K_{nnmm}=0$. Taking the symmetries $L_{abcd}=L_{cdab}$, $D_{abcd}=D_{cdab}$ into account, we get from (39):
\begin{equation}
\frac{d\rho_{nn}(t)}{dt}=-\int_{\infty}^{t}dt_1e^{\epsilon(t-t_1)}\sum_{m\neq n}[K_{nnmm}(t-t_1)\rho_{mm}(t_1)-K_{mmnn}(t-t_1)\rho_{nn}(t_1)].
\end{equation}
Next we shall make the following approximations. With accuracy up to the value of $(\Delta x_0)^2$ order in the exponent in expression (40) we shall replace the complete Liouville operator $\hat{L}=\hat{L}_0+\hat{L}'$ with $\hat{L}_0$. With the same precision we may set $\rho_{nn}(t-t')=\rho_{nn}(t)$. As a result, from (43) we arrive at
\begin{equation}
\frac{d\rho_{nn}(t)}{dt}=\sum_{m\neq n}[W_{nm}\rho_{mm}(t)-W_{mn}\rho_{nn}(t)],
\end{equation}
where
\begin{equation}
W_{nm}=-\int_{-\infty}^{0}dt' e^{\epsilon t'}[\hat{L}' e^{it(1-\hat{D})\hat{L}_0}(1-\hat{D})\hat{L}']_{nnmm}.
\end{equation}
Since
\begin{equation}
(\hat{D}\hat{L}_0)_{abcd}=\delta_{ab}[\phi_a\delta_{ac}\delta_{bd}-\phi_b\delta_{bd}\delta_{ac}]=0,
\end{equation}
in expression (45) the $\hat{D}$ operator in the argument of exponential function may be omitted.

Taking into account time dependence of the operator
$x(t)=x_o+\Delta
x_of(t),~f(t)=\sum\limits_{\nu=-\infty}^{\infty}e^{i\nu\Omega
t},~\Omega=2\pi/T_o$  for
non-diagonal matrix elements we get
\begin{equation}
\frac{d\rho _{nn}(t)}{dt}=\sum_{m\not
=n}(W_{nm}\rho_{mm}(t)-W_{mn}\rho_{nn}(t)),
\end{equation}
where  $W_{nm}=\frac{\pi}{2}
|V_{nm}|^2\sum\limits_{\nu=-\infty}^{\infty} \delta
(E_{nm}-\nu\Omega)$ is the transition amplitude between the
eigenstates of the Hamiltonian $\hat{H}_o~$ (3), $E_{nm}=\phi_n-\phi_m$
, $V_{nm}$ is the matrix element of the operator $\hat{V}_o=\Delta
x_oQ^2_1Q^2_2$ in the basis of eigenfunctions of the Hamiltonian
$\hat{H}_o,~ V_{nm}=\langle\psi_n|\hat{V}_o|\psi_m\rangle.$

Equation (47) describes non-reversible evolution of the system
from non-stationary state to the stationary state defined by the
principle of detail equilibrium [20]. To prove irreversibility of
the process let us consider time dependence of non-equilibrium
entropy [30]

\begin{equation}
S(t)=-K_B\sum_n\rho_{nn}(t)\ln(\rho_{nn}(t)),\label{25}
\end{equation}
where $K_B$ is the Boltzmann  constant. Taking into account
$\sum_n\rho_{nn}(t)=1$ , from (48) we get
$$
\frac{dS(t)}{dt}=-K_B\sum_n\sum_mW_{nm}[\rho_{mm}(t)-\rho_{nn}(t)]\ln(\rho_{nn}(t))-
$$
\begin{equation}
-K_B\sum_n\frac{\partial \rho_{nn}(t)}{\partial
t}=\frac{1}{2}K_B\sum_n\sum_mW_{nm}[\rho_{nn}(t)-\rho_{mm}(t)][\ln(\rho_{nn}(t))-\ln(\rho_{mm}(t))].
\end{equation}

Due to the property of logarithmic function
$$(\rho_{nn}(t)-\rho_{mm}(t))(\ln(\rho_{nn}(t))-\ln(\rho_{mm}(t)))\geq
0$$ we see that $\frac{dS}{dt}\geq 0$. This testifies the growth
of entropy during the evolution process.

More exact estimation of entropy growth may be obtained from the
principle of detail equilibrium [20]. It is evident from the
expression of transition probability  $W_{nm}=\frac{\pi}{2}
|V_{nm}|^2\sum\limits_{\nu=-\infty}^{\infty} \delta
(E_{nm}-\nu\Omega)$ that only resonant states $E_{nm}=\nu \Omega$
are included in the process. Transitions between states lead to
the redistribution of the initial probabilities given by
$\rho_{nn}(t_o)$. As a result after the time being we have
stationary distribution of the probabilities defined by the
principle of detail equilibrium.

According to this principle number of transitions from the state
$m$ into the state $n$ is equal to the quantity of reverse
transitions
\begin{equation}
W_{nm}\rho_{mm}=W_{mn}\rho_{nn}.
\end{equation}

Taking into account that in our case $W_{nm}=W_{mn}$ , for the
entropy growth we get
\begin{equation}
\Delta S=S(t\gg \sqrt{2}/\sigma)-S(t=0)=K_B\ln N,
\end{equation}
where the number of levels $N$, completely included into the process
is defined by the spectral characteristics of the
Hamiltonian (3) and perturbation (according to (47) ).

The authors express their gratitude to Professor A.Ugulava for
valuable suggestions and useful discussions.

\end{document}